\documentclass[%
 amsmath,amssymb,
 aps,
]{revtex4-2}

\usepackage{graphicx}
\usepackage{dcolumn}
\usepackage{bm}
\usepackage{tabularx}
\usepackage{subcaption}
\usepackage{booktabs}
\usepackage{xcolor}

\newcommand{\PP}{\mathbb{P}}

\newcommand{\col}{{\rm col}}

\newcommand{\TT}{{\mathbb{T}}}

\begin{document}


\title{From Ising to Potts: Physics-inspired Potts machines of coupled oscillators for low-energy sampling and combinatorial optimization}

\author{Yi Cheng}
 \altaffiliation[]{zss7gw@virginia.edu}
\author{Zongli Lin}%
 \email{zl5y@virginia.edu}
\affiliation{%
Charles L. Brown Department of Electrical and Computer Engineering, University of Virginia, Charlottesville, Virginia 22904, USA.
}%




\date{\today}

\begin{abstract}
The $q$-state Potts model is a fundamental model in statistical physics that generalizes the Ising model and plays a key role in the study of phase transitions, critical phenomena, complex systems, and combinatorial optimization. Sampling low-energy configurations of the $q$-state Potts model is essential to these studies, but it remains challenging. While physics-inspired dynamical sampling has been extensively explored for the Ising case ($q=2$) in the form of Ising machines, its generalization to general $q$-state Potts models remains largely unexplored. To fill this gap, we propose a class of physics-inspired dynamical samplers that directly target general
$q$-state Potts models, which we refer to as the oscillator Potts machine (OPM). We show, through theoretical analysis and numerical experiments, that the OPM exhibits a systematic low-energy bias with respect to the underlying Potts energy landscape. Furthermore, we demonstrate, via phase perturbation analysis, that the OPM, as overdamped Langevin dynamics, can be realized with a network of self-sustaining oscillators, demonstrating that the OPM is naturally realizable in hardware using standard technology such as CMOS. We design a small-scale ring-oscillator circuit that implements a three-state OPM and validate its operation through transistor-level simulation. Leveraging the low-energy bias of the OPM for Potts models, we then apply it to large-scale max-$K$-cut problems by mapping these instances to $q$-state Potts Hamiltonians and compare its performance against established algorithms. Our results position the OPM as a promising, physically grounded dynamical system framework for multi-state sampling and combinatorial optimization.
\end{abstract}

\maketitle


\section{Introduction}
The $q$-state Potts model is a canonical model in statistical physics that generalizes the Ising model and arises in diverse applications, from combinatorial optimization \cite{basak2017universality},  neuroscience \cite{kanter1988potts}, to inference \cite{peng2003neuron}. Efficiently sampling its low-energy configuration is central to these applications, but it remains notoriously challenging. As noted in \cite{blanca2024tractability}, on a general graph, this problem may be computationally hard, and the hardness holds at arbitrarily low temperatures.

Recently, rapid advances in hardware platforms have spurred growing interest in physics-inspired dynamical samplers, which exploit continuous-time, massively parallel dynamics to explore rugged energy landscapes. Examples include analog and digital-analog implementations of Ising machines based on optical parametric oscillators \cite{ng2022efficient,inagaki2016coherent,mcmahon2016fully, honjo2021100,yamamoto2017coherent,cen2022large,wang2025efficient,sakellariou2025encoding}, CMOS and mixed-signal circuits \cite{bohm2022noise,chou2019analog,mohseni2022ising,lee2025noise,maher2024cmos,chen2024oscillatory,wang2021solving,graber2024integrated, deng2024mems}, and simulated bifurcation dynamics \cite{goto2019combinatorial,kanao2022simulated,tatsumura2019fpga}. These systems are designed so that their physical or virtual dynamics approximately follow the gradient of an Ising energy function, often with added noise or bifurcation mechanisms, thereby realizing energy-based sampling or approximate optimization in hardware with competitive power and speed performance compared to digital implementations. The success of these Ising-type dynamical samplers in solving large-scale quadratic unconstrained binary optimization \cite{singh2024uplink,zheng2025constrained} and Ising benchmark problems \cite{mcmahon2016fully,inagaki2016coherent} has highlighted the promise of physics-inspired approaches to combinatorial optimization and probabilistic inference.

Despite this progress, most physics-inspired dynamical samplers have been confined to the Ising case ($q=2$). Extending them to general multi-state Potts models is nontrivial. A straightforward encoding of a $q$-state variable into multiple Ising spins \cite{inaba2022potts,kawakami2023constrained,shukla2025non,garg2025efficient,gonul2025multi} typically introduces auxiliary variables with additional constraints and dense couplings, which inflate hardware resource requirements and may distort the original energy landscape. In addition to such encoding-based approaches, oscillator-based dynamical models have recently been proposed that aim to address max-$K$-cut and related problems by directly engineering multi-phase coupling functions and external injections in coupled oscillator networks \cite{mallick2022computational}. In these formulations, the effective interaction is realized through carefully designed, problem-dependent phase response functions that are expressed as singular limits of Gaussian mixtures, leading to relatively intricate phase dynamics, whose physical interpretation and relation to a simple, canonical Potts Hamiltonian are less transparent. While these works represent important steps toward a multi-state, physics-inspired sampler, a general and structurally simple dynamical sampler that directly targets the $q$-state Potts model, with a clear energy function and hardware-friendly structure, is still lacking.

\begin{figure}[htbp]
    \centering
    \includegraphics[width=1\linewidth]{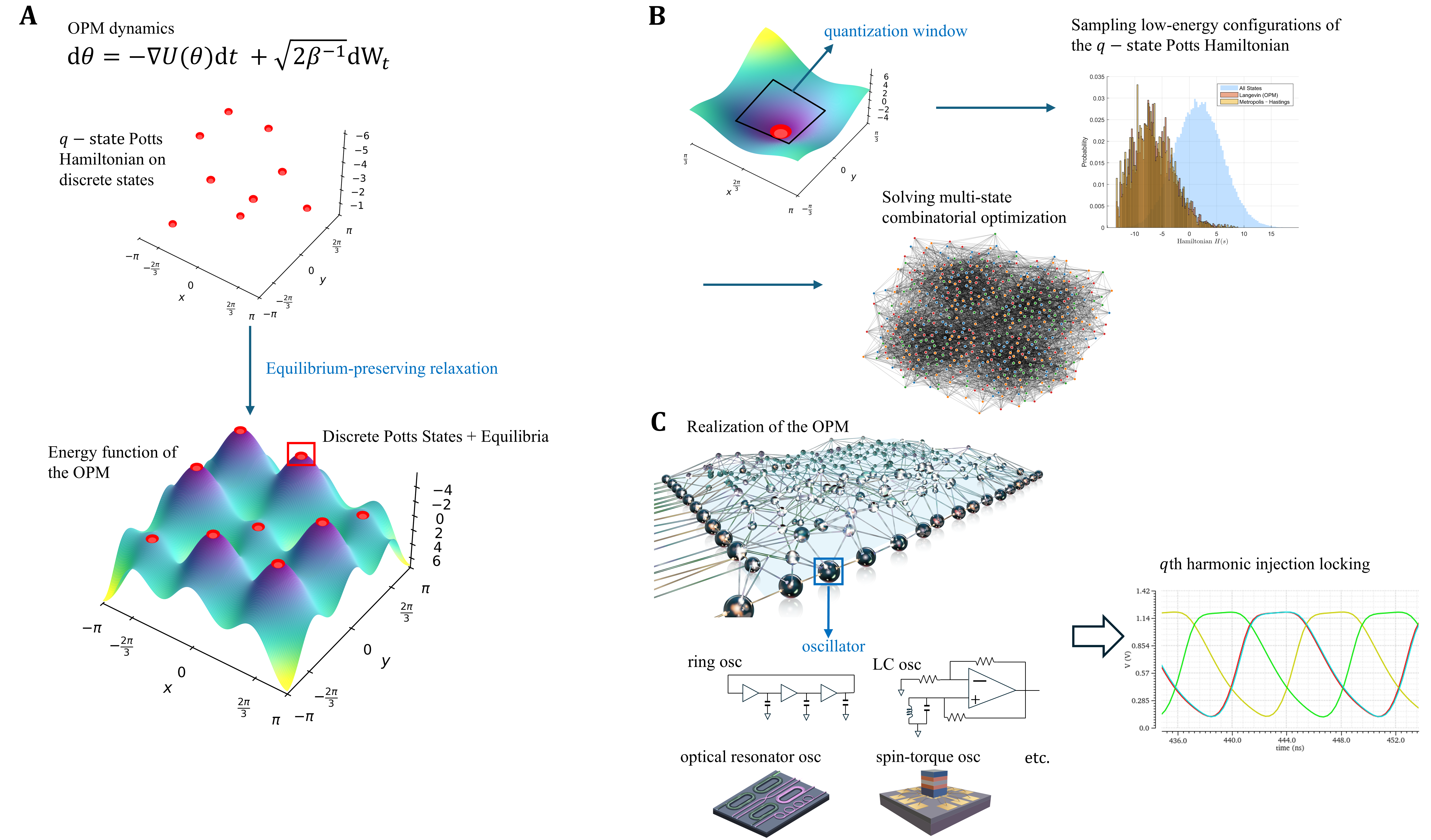}
    \caption{Overview of the OPM: (A) The energy function of an OPM associated with a $q$-state Potts Hamiltonian is an equilibrium-preserving relaxation that enables overdamped Langevin dynamics (gradient flow with noise) over continuous phases. Equilibrium-preserving relaxation means the continuous energy $U(\theta)$ is constructed so that every discrete Potts configuration corresponds to a structurally stable equilibrium (``sampling point'') of the continuous dynamics, and $U(\theta)$ evaluated at those points reproduces the Potts Hamiltonian (up to constants/rescaling). (B) Continuous phases are mapped back to discrete Potts states via a quantization window, yielding a discrete sampling distribution that remains biased toward low-energy configurations. (C) The dynamics can be realized with a coupled oscillator network using $q$th-harmonic injection locking to produce multi-stable phase states compatible with hardware implementation.}
    \label{Introduction}
\end{figure}

In this work, we fill the gap mentioned above by introducing a continuous, physics-inspired dynamical system that directly targets the $q$-state Potts model. We refer to this system as the oscillator Potts machine (OPM). The OPM is built as overdamped Langevin dynamics, whose energy function acts as an equilibrium-preserving relaxation of the Potts Hamiltonian (see Fig.~\ref{Introduction}A). We further prove that, after introducing a natural quantized readout of the continuous phases, the resulting discrete distribution retains a provable low-energy bias with respect to the underlying Potts energy landscape over a finite range of quantization windows (see Fig.~\ref{Introduction}B). Next, we demonstrate, via phase perturbation analysis, that the OPM dynamics can be realized with a network of self-sustaining oscillators, such as a CMOS ring oscillator network, establishing a clear path from the abstract dynamical sampler to concrete hardware realizations (see Fig.~\ref{Introduction}C).  Finally, we use the OPM to solve max-$K$-cut problems (for $K=3,4,5$) on standard Gset (G1–G5) benchmarks to demonstrate its effectiveness in solving large-scale combinatorial optimization problems. These properties position the OPM as a scalable, physically grounded framework for direct dynamical sampling and approximate optimization of general
$q$-state Potts models.

\section{Results}
\subsection{From the $q$-state Potts model to the OPM}\label{OPMintro}
\begin{figure}[htbp]
    \centering
    \includegraphics[width=1\linewidth]{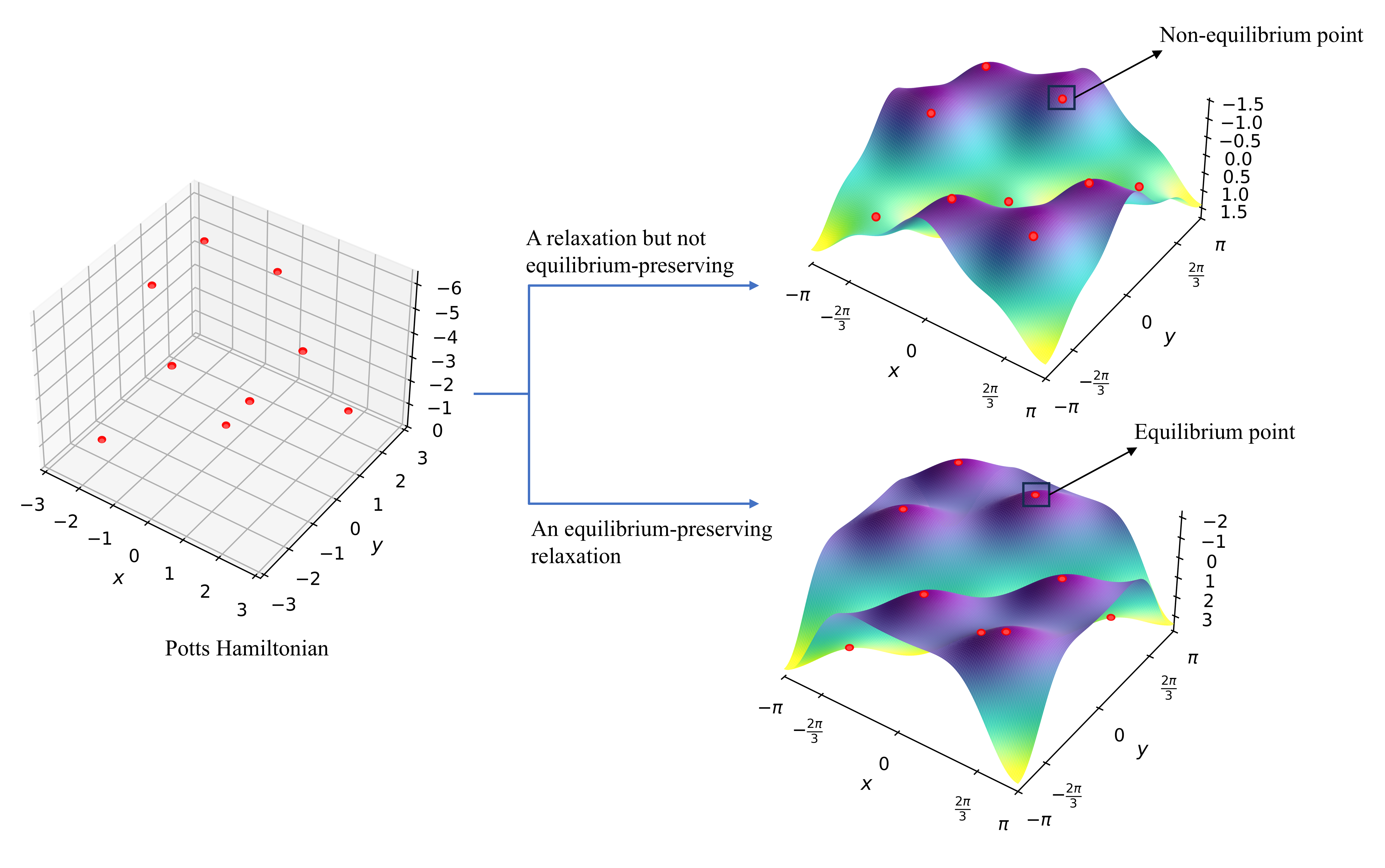}
    \caption{Two different relaxations of a $q$-state Potts Hamiltonian: the non-equilibrium-preserving relaxation (upper) and the equilibrium-preserving relaxation (lower).}
    \label{equilibriumRXvsRx}
\end{figure}

The $q$-state Potts model is a generalization of the Ising model and allows each spin to take $q\ge 2$ arbitrary distinct values, say, without loss of generality, $s_i\in \{0,1,\ldots,q-1\}$, where $s_i$ denotes the state of the $i$th spin for $i=1,2,\ldots,N$. The collective behavior of these spins is described by the Potts Hamiltonian
\begin{eqnarray} \label{PottsHamil}
    H_{\rm Potts}(s) = -\sum_{i<j}J_{ij}\delta(s_i,s_j),\; \delta(s_i,s_j) =\begin{cases}
    1, \mbox{ if } s_i=s_j,\\
    0, \mbox{ if } s_i\ne s_j,
\end{cases}
\end{eqnarray}
where $s = \col\{s_1,s_2,\ldots,s_N\}$ is the spin configuration and $J_{ij}$ denotes the coupling weight between spins $i$ and $j$. The probability of observing a spin configuration obeys the Boltzmann distribution
\begin{equation} \label{boltzman}
    P(s) = \frac{1}{Z}{\rm e}^{-\beta H_{\rm Potts}(s)}, \; Z = \sum_{{\rm All}\ s}{\rm e}^{-\beta H_{\rm Potts}(s)},
\end{equation}
where $\beta$ is the inverse temperature and $Z$ is the partition function ensuring normalization. The aim of this work is to study a class of physics-inspired dynamical systems that directly samples the low-energy configuration of the $q$-state Potts model. The energy function of the system should be an equilibrium-preserving relaxation of the Potts Hamiltonian \eqref{PottsHamil}. Here, ``equilibrium-preserving relaxation'' requires not only that the system's energy function be a relaxation of the Potts Hamiltonian at certain points, but also that all these points be structurally stable equilibrium points of the energy function. Points where the gradient of the energy function vanishes are commonly referred to as stationary points. In the context of overdamped Langevin dynamics, these stationary points coincide with the equilibrium points of the deterministic drift. For convenience of reference, throughout this paper we refer to stationary points as equilibrium points. We recall that ``a structurally stable equilibrium point'' is an equilibrium point that remains an equilibrium point regardless of values of parameter configurations of the energy \cite{cheng2024control}. Fig.~\ref{equilibriumRXvsRx} illustrates the difference between the non-equilibrium-preserving relaxation and the equilibrium-preserving relaxation.
This property is particularly important for physics-inspired samplers.
In a generic relaxation, the target Potts configurations may agree with the relaxed energy at certain points of the energy function in value only, but they need not be mapped to equilibrium points of the underlying continuous dynamics. As a consequence, the dynamical sampler may spend most of its time near other spurious equilibrium points that have no direct counterpart in the original Potts model. By contrast, an equilibrium-preserving relaxation guarantees that the candidate Potts configurations constitute actual equilibrium (typically metastable) states of the dynamics. In the presence of noise, the sampler will wander between these physically meaningful states, and the structure of their basins of attraction and energy barriers is directly inherited from the Potts Hamiltonian. This not only improves the interpretability and robustness of hardware implementations but also provides a more faithful starting point for analyzing the sampling bias relative to the target distribution. We thereby propose the OPM with the following dynamics,
\begin{eqnarray}\label{OPMdyn}
  {\rm d}[\theta_t]_i = -\! \left\{ \rule{0cm}{.6cm}\! K\! \sum_{j=1}^N \!J_{ij}\!\! \left(\sum_{m=1}^{q-1}\!m(q-m)\sin\big(m([\theta_t]_i\!-\![\theta_t]_j)\big)\!\!\right)\right. \left.+K_{\rm s}\sin(q[\theta_t]_i)\rule{0cm}{.6cm}\right\}{\rm d}t + \sqrt{2\beta^{-1}} {\rm d}W_t,
\end{eqnarray}
where $[\theta_t]_i \in \TT:=[0,2\pi)$ is the random variable of the $i$th oscillator, $K$ and $K_{\rm s}$ are two positive tunable parameters, and ${\rm d}W_t$ is the Wiener process. The energy function of the OPM can be verified to be
\begin{eqnarray}\label{Utheta}
    U(\theta) = -\frac{K}{2}\sum_{i<j}J_{ij} \left(2\sum_{m=1}^{q-1}(q-m)\cos(m(\theta_i-\theta_j))\right) -\frac{K_{\rm s}}{q}\sum_{i=1}^N\cos(q\theta_i)
\end{eqnarray}
such that ${\rm d}{ U(\theta)}/{\rm d}t \leq 0$ along the trajectory of \eqref{OPMdyn} when the stochastic part is omitted. Before we discuss further details of the OPM, we observe that the commonly studied oscillator Ising machine is a special case of \eqref{OPMdyn} with $q=2$. From a computational standpoint, the OPM dynamics have two notable features. First, the governing equations take the form of stochastic differential equations (SDEs), which can be directly integrated numerically on conventional digital computers, naturally leveraging modern graphics processing unit (GPU) acceleration, without requiring any specially designed hardware devices or custom circuits. Second, the same dynamics can also, in principle, be mapped onto a broad class of self-sustaining oscillator networks. In particular, in Section \ref{physicalrealization}  we demonstrate how to realize them with a ring-oscillator-based circuit, suggesting that, using technologies such as CMOS, the OPM dynamics have the potential to enable low-power, high-performance, large-scale integrated computing architectures.

\subsection{Dynamical properties of the OPM}\label{dynamicalpro}

To better show the dynamical properties of the OPM, we temporarily ignore the Wiener process in \eqref{OPMdyn}. In the absence of the Wiener process, the dynamics \eqref{OPMdyn} simplifies to
\begin{equation}\label{dyn_OPM_st}
    {\dot \theta}_i = - K \sum_{j=1}^N J_{ij} \left(\sum_{m=1}^{q-1}m(q-m)\sin(m(\theta_i-\theta_j))\right)-K_{\rm s}\sin(q\theta_i).
\end{equation}
We have previously mentioned that the energy function of the OPM is an equilibrium-preserving relaxation of the Potts Hamiltonian. We now explain why this is the case. Recall that equilibrium-preserving relaxation is a more restricted relaxation that imposes an additional requirement on the energy function, that is, certain points of the energy function of the OPM that have been the relaxation of the Potts Hamiltonian should also be equilibrium points of the energy function. Thus, we need to explain it from the following two aspects. First, the energy function of the OPM possesses certain structurally stable equilibrium points. Second, the energy function of the OPM is indeed a relaxation of the Potts Hamiltonian at these equilibrium points.

From a dynamical-system perspective, the OPM dynamics can, in principle, admit a large number of equilibrium points. Among them, the class that is most special and relevant for computation is the structurally stable equilibrium points, namely those that remain equilibrium points under any perturbation of the parameters $J_{ij}$, $K$, and $K_{\rm s}$. Our analysis shows that any phase vector $\theta^\star = \{\theta_1^\star,\theta_2^\star,\ldots,\theta_N^\star\}$ with components $\theta_i^\star \in \{2\pi k/q:k=0,1,\ldots, q-1\}$ is a structurally stable equilibrium point of the OPM dynamics. For convenience of reference, we refer to such equilibrium points as sampling points. There is a natural one-to-one correspondence between sampling points and Potts spin configurations via mapping $s[\theta^\star] : \{2\pi k/q:k=0,1,\ldots,q-1\}^N \to \{0,1,\ldots,q-1\}^N$,  defined componentwise by $s_i[\theta^\star] = k$ if $\theta_i^\star = 2\pi k/q$, $k=0,1,\ldots,q-1$. With this mapping, we can show that the OPM energy evaluated at a sampling point $\theta^\star$ is related to the Potts Hamiltonian evaluated at the corresponding spin configuration $s[\theta^\star]$ by
\begin{equation}\label{UH}
    U(\theta^\star) = \frac{K\big(q(q-1)+q\big)}{2}H_{\rm Potts}(s[\theta^\star]) + \frac{K}{2}\sum_{i<j}J_{ij}q - \frac{K_{\rm s}N}{q}.
\end{equation}
Equation \eqref{UH} makes it clear that, when restricted to the sampling points, the OPM energy is indeed a relaxation of the Potts Hamiltonian. Together with the fact that all Potts configurations correspond to structurally stable equilibrium points of the OPM dynamics, this shows that the OPM energy defines an equilibrium-preserving relaxation of the Potts Hamiltonian in the sense of the definition above. The proofs of the above results are provided in the Supplementary Materials.

As mentioned above, besides the sampling points, the OPM dynamics also admit other types of equilibrium points, which we collectively refer to as non-sampling points. From the standpoint of sampling, these non-sampling points are undesirable since trajectories can become trapped at such equilibrium points or spend an excessive amount of time wandering in their vicinity, thereby reducing the efficiency with which the OPM visits the sampling points. Our analysis shows that, when the ratio $K_{\rm s}/K$ is sufficiently large, all non-sampling points are unstable, whereas all sampling points are asymptotically stable. This result implies that, by appropriately tuning the parameters $K_{\rm s}$ and $K$, the detrimental influence of non-sampling points can be effectively suppressed. Moreover, we prove that, as $K_{\rm s}/K \to \infty$, the domains of attraction of all sampling points converge to congruent hypercubes in the phase space, each centered at a sampling point and having the same size. Again, all the above analyses are provided in the Supplementary Materials.

It is also worth emphasizing that, from a dynamical system point of view, the OPM is fundamentally different from binary-encoding-like approaches to Potts models \cite{wang2021solving,inaba2022potts,kawakami2023constrained,shukla2025non,garg2025efficient}. In binary encodings, a single
$q$-state Potts spin is represented by a collection of binary variables (e.g., one-hot encodings), so that realizing an $N$-spin $q$-state Potts model requires an expanded network with many more physical nodes and coupling weights. In contrast, each OPM oscillator naturally supports $q$ distinct stable phase states, and the Potts configuration is encoded directly in these phases. Consequently, an $N$-spin $q$-state Potts model can be implemented with $N$ oscillators and a single set of pairwise couplings $\{J_{ij}\}$, without any duplication of oscillators or couplings in an enlarged binary space. This difference has important implications for both hardware implementation and numerical simulation. On the hardware side, avoiding an explicit binary encoding significantly reduces the required number of oscillators, interconnections, and control lines, which translates into a smaller size, a lower power consumption, and a simpler physical layout. On the algorithmic side, simulating the OPM dynamics only involves integrating the original $N$-oscillator system, rather than a higher-dimensional binary surrogate, leading to substantial savings in computational cost and memory footprint. Table~\ref{tab:graph_opm_oim} summarizes the resource requirements of the OPM and a representative Ising scheme \cite{wang2021solving} for sampling the same  $q$-state Potts model, highlighting the potential advantages of the inherent multi-state representation of the OPM.
\begin{table}[htb]
\centering
\renewcommand{\arraystretch}{0.8} 
\setlength{\tabcolsep}{5pt}      

\begin{tabular}{l c c c}
\toprule
 & $\#$ \textbf{Nodes (oscillators)} & $\#$ \textbf{Edges (couplings)} & $\#$ \textbf{Solution space} \\
\midrule
The original problem  & $N$ & $e$ & $2^{N\log_2q}$\\
Oscillator Potts machine & $N$ & $e$ & $2^{N\log_2q}$ \\
Oscillator Ising machine & $qN+1$ $\uparrow$ & $qe+qN + Nq(q-1)/2$  $\uparrow$  & $2^{Nq}$  $\uparrow$  \\
\bottomrule
\end{tabular}
\smallskip
\caption{Comparison between the original problems, the OPM, and the Ising machine.}
\label{tab:graph_opm_oim}
\end{table}

\subsection{Low-energy bias after quantization}\label{lowenergybias}
Since the energy function of the OPM is given by equation \eqref{Utheta}, the dynamics \eqref{OPMdyn} can be written as
\begin{equation}\label{Lagevin}
   {\rm d}\theta_t = - \nabla U(\theta_t){\rm d}t +\sqrt{2\beta^{-1}}{\rm d}W_t,
\end{equation}
which is the standard form of overdamped Langevin dynamics. Since the energy function $U(\theta)$ is smooth in $\TT^N$ and $\rm e^{-\beta U(\theta)}$ is Lebesgue integrable in $\TT^N$, the distribution of \eqref{Lagevin} converges to the distribution with probability density function $\Pi(\theta) = Z_{\rm L}^{-1}{\rm e}^{-\beta U(\theta)}$, where $Z_{\rm L} = \int_{\theta \in \TT^N}{\rm e}^{-\beta U(\theta)}{\rm d}\theta$ \cite{leimkuhler2016computation,roussel2018spectral}. In using the OPM dynamics \eqref{Lagevin} as a sampler, however, our ultimate goal is not to sample arbitrary continuous phase configurations, but rather to sample a finite set of discrete equilibrium configurations (e.g., the sampling points). This creates a conceptual gap between the continuous Langevin description and the discrete sampling task, that is, the function $\Pi(\theta)$ is the density function  with respect to $\theta$, so the probability assigned to any individual point, say $\theta^\star$, in $\TT^N$ is given by $\PP(\theta^\star) = \int_{\theta = \theta^\star}\Pi(\theta) {\rm d}\theta = 0$. In other words, $\Pi(\theta)$ does not directly provide a probability mass for each discrete equilibrium configuration we care about.

In practice, every implementation of the OPM,  whether in numerical simulation or in hardware, includes an implicit quantization step that maps a continuous phase vector $\theta_t$ to a discrete configuration $\theta^\star$. To make this explicit, let $Q : \TT^N \to \mathcal{S}$ denote a quantization map from the continuous state space to a finite set $\mathcal{S}$ of candidate discrete configurations. The probability that the OPM sampler outputs a configuration $\theta^\star \in \mathcal{S}$ is then given by the $\Pi$-measure of its quantization cell $\PP_{\rm q}(\theta^\star)= \int_{\theta_t \in Q^{-1}(\theta^\star)}\Pi(\theta_t){\rm d}\theta_t$. In particular, the quantization map $Q$ we consider is such that a sample $\theta_t$ drawn from \eqref{Lagevin} falling within a hypercube of side $2a$ centered at a sampling point $\theta^\star$, $\left(-a,+a\right)_{\theta^\star}^N$, is assigned the value of $\theta^\star$, that is, $Q^{-1}(\theta^\star) = (-a,+a)_{\theta^\star}^N$. Then, the probability of $\theta^\star$ occurring after the quantization is
\begin{eqnarray}\label{distributionSP}
    \PP_{\rm q}(\theta^\star) && =\int_{\theta_t \in Q^{-1}(\theta^\star)}\Pi(\theta_t){\rm d}\theta_t=Z_{\rm L}^{-1}\int_{\theta_t \in \left(-a,+a\right)_{\theta^\star}^N}{\rm e}^{-\beta U(\theta_t)} {\rm d}\theta_t.
\end{eqnarray}
Introducing this quantization framework bridges the gap between the continuous Langevin model and the discrete sampler and will allow us to analyze the low-energy bias induced by quantization in the following discussion. We note that quantization is very common in many physics-inspired systems for solving combinatorial optimization problems \cite{goto2019combinatorial,zhang2022review,wang2023bifurcation}, since the internal states of physics-inspired systems evolve continuously due to their analog nature, whereas solving combinatorial optimization problems inherently requires discrete solutions.

Clearly, for any two sampling points $\theta^\star$ and $\phi^\star$, $\PP_{\rm q}(\theta^\star)/\PP_{\rm q}(\phi^\star) \to {\rm e}^{\beta(U(\phi^\star)-U(\theta^\star))}$ as $a \to 0$. On the one hand, as $a \to 0$, although the ratio of probabilities between any two sampling points remains exactly the one prescribed by the Boltzmann distribution \eqref{boltzman}, the efficiency of discrete sampling becomes extremely low. On the other hand, for larger values of $a$, while the efficiency of discrete sampling is significantly improved, the resulting distribution \eqref{distributionSP} no longer preserves the crucial monotonicity property, that is, $U(\theta^\star)>U(\phi^\star)$ no longer theoretically guarantees $\PP_{\rm q}(\theta^\star)<\PP_{\rm q}(\phi^\star)$. Thus, there should be a trade-off between the monotonicity property of equation \eqref{distributionSP} and the efficiency of discrete sampling by evaluating the value of $a$. We further provide an analytical result showing that, as long as the parameter $a$ lies in a suitable interval, the distribution in \eqref{distributionSP} still preserves the desired monotonicity property, that is, if $U(\theta^\star)>U(\phi^\star)$, then $\PP_{\rm q}(\theta^\star)<\PP_{\rm q}(\phi^\star)$ (see Supplementary Materials). From a theoretical perspective, this establishes a first qualitative characterization of the low-energy bias induced by the OPM dynamics and thus offers a justification for using OPM as a sampler to sample low-energy configurations and to solve combinatorial optimization. It should be emphasized, however, that this monotonicity guaranty is a rather conservative property. In particular, it does not quantify how close the ratio $\PP_{\rm q}(\theta^\star)/\PP_{\rm q}(\phi^\star)$ is to the ideal Boltzmann factor ${\rm e}^{-\beta (U(\theta^\star)-U(\phi^\star))}$, nor does it reveal whether \eqref{distributionSP} enjoys stronger properties beyond mere monotonicity. Deriving sharper, possibly quantitative, guarantees on how well $\PP_{\rm q}$ approximates the Boltzmann distribution, such as bounds on the deviation of these probability ratios, remains an interesting direction for future work.

\subsection{Physical realization of the OPM with self-sustaining oscillator networks}\label{physicalrealization}

To physically implement the $q$-state OPM, we exploit the fact that the deterministic OPM dynamics \eqref{dyn_OPM_st} can be mapped onto the phase dynamics of weakly coupled self-sustaining oscillators. Consider an autonomous oscillator with a natural frequency $f$ and a steady-state waveform $x_{\rm s}(t) = x(ft)$. Under a small time-varying perturbation $b(t)$,  perturbation projection vector (PPV) theory describes the oscillator’s response as $x(ft+f\alpha(t))$, where the time shift $\alpha(t)$ obeys $\dot \alpha(t) = p(ft+f\alpha(t))b(t)$ with a $1$-periodic PPV $p(t)$ \cite{demir1998phase,levantino2012computing,neogy2012analysis}. Extending this description to $N$ identical oscillators yields $N$ time-shift equations $\dot \alpha_i(t) = p(ft+f\alpha_i(t))b_i(t)$, where $b_i(t)$ represents the external inputs for the $i$th oscillator.

\begin{figure}[htbp]
    \centering
    \includegraphics[width=0.9\linewidth]{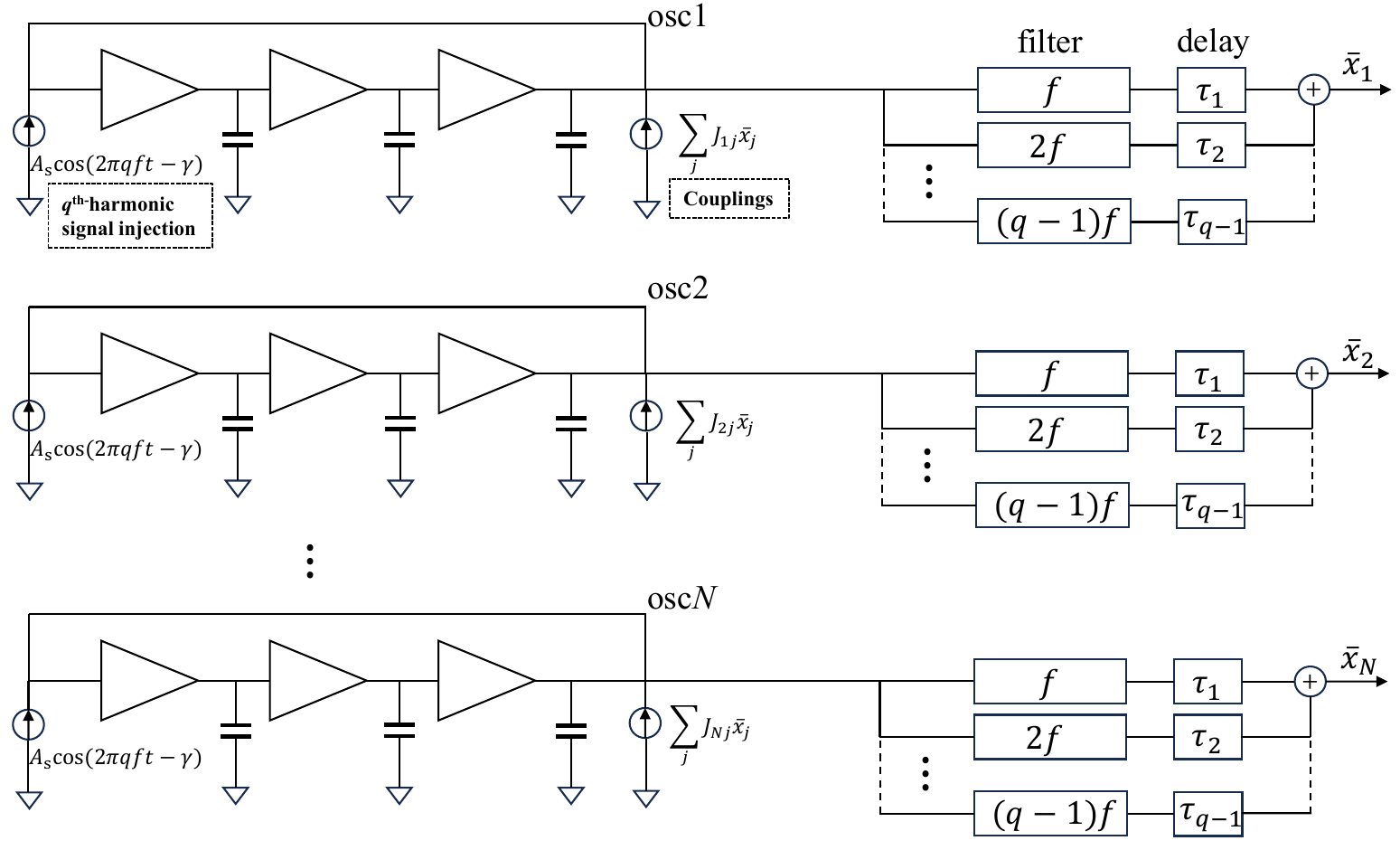}
    \caption{A schematic diagram of the circuit design of the OPM. Example with a three-stage ring oscillator (applicable to any odd-stage ring oscillator). There are $q-1$ band-pass filters with center frequencies $f,2f,\ldots,(q-1)f$ and $q-1$ tunable delay components for each oscillator.}
    \label{circuitOPM}
\end{figure}

For each oscillator, we attach $q-1$ harmonic channels, each consisting of a band-pass filter and a tunable delay line, whose center frequencies are set to $f,2f,\ldots,(q-1)f$ and whose tunable time delays are set to $\tau_1,\tau_2,\ldots,\tau_{q-1}$, respectively. When the oscillator output passes through the $m$th channel, the resulting signal is approximately $A_m|a_m|\cos(2\pi fm(t+\alpha_i)+\phi_m-\varphi_m-\tau_m)$, where $|a_m|$ and $\phi_m$ are the magnitude and phase of the $m$th Fourier coefficient of the injected node waveform, $\varphi_m$ is the phase shift introduced by the filter, and $A_m$ and $\tau_m$ are tunable amplitude and delay. Summing over the $q-1$ channels produces a synthesized coupling signal $\bar x_i$, and the external input to the $i$th oscillator is chosen as a weighted sum of other oscillators’ synthesized coupling signals plus a global $q$th harmonic injection, that is, $b_i(t) = \sum_{j=1}^NJ_{ij}\bar x_j(ft+f\alpha_i(t)) + A_{\rm s}\cos(2\pi q ft-\gamma)$, where $A_{\rm s}$ and $\gamma$ are tunable amplitude and time delay. Under the standard assumptions of weak coupling and slowly varying phases, we average the equations over one period and obtain
\begin{eqnarray}\label{fastaverage3}
    \dot \alpha_i(t) &\approx&  \sum_{j=1}^N J_{ij}\left(\sum_{m=1}^{q-1} A_m|a_m||p_m| \cos(2\pi f m(\alpha_i(t)-\alpha_j(t))-\varphi_m+\phi_m+\tau_m+\theta_m)\right) \nonumber \\
    &&+ A_{\rm s}|p_q| \cos(2\pi f q\alpha_i(t)+\gamma +\theta_q),
\end{eqnarray}
where $|p_m|$ and $\theta_m$ are the magnitude and phase of the $m$th Fourier coefficient of the PPV waveform. Note that $\varphi_m$, $\phi_m$, $\theta_m$, $\theta_q$, $|a_m|$, $|p_m|$, and $|p_q|$ are parameters to be measured, while $\tau_m$, $\gamma$, $A_m$, and $A_{\rm s}$ are tunable parameters. By appropriately choosing $A_m,\tau_m$ and the amplitude and phase of the $q$th harmonic injection, the averaged equation \eqref{fastaverage3} reduces to
\begin{equation*}
    \dot \alpha_i(t) \approx \sum_{j=1}^N J_{ij}\left(\sum_{m=1}^{q-1}m(q-m)\sin(2\pi f m (\alpha_i(t)-\alpha_j(t)))\right) - A_{\rm s}|p_q| \sin(2\pi f q \alpha_i(t)),
\end{equation*}
or an equivalent form with an opposite sign convention for the coupling term, depending on how $J_{ij}$ is defined in the circuit. After a linear rescaling of variables $\theta_i=2\pi f \alpha_i$, this averaged phase model becomes identical in structure to the deterministic part of the OPM dynamics introduced in \eqref{dyn_OPM_st}, with the OPM parameters $K$ and $K_{\rm s}$ expressed in terms of the circuit-level quantities $A_m$, $A_{\rm s}$, $|a_m|$, and $|p_m|$. In this way, OPMs can be realized by a broad class of oscillator networks, with oscillator-based Ising machines recovered as the special case $q=2$ and $m=1$. A schematic diagram of the circuit design of the OPM is shown in Fig.~\ref{circuitOPM}.
We note that, in Refs \cite{wang2017oscillator,roychowdhury2022oscillator}, the derivation and implementation of the oscillator Ising machine and the clock machine are also carried out with networks of self-sustaining oscillators, where the oscillator coupling involves only first-order interaction terms ($m=1$). In the present work, we follow this line of research and generalize implementations that accommodate higher-order coupling terms. This formulation can be viewed as treating oscillator Ising machines as a special case within the OPM framework and suggests a possible route for generalizing binary Hamiltonian samplers to their multi-valued counterparts.

\begin{figure}[htbp]
     \centering
    \includegraphics[width=1.5in]{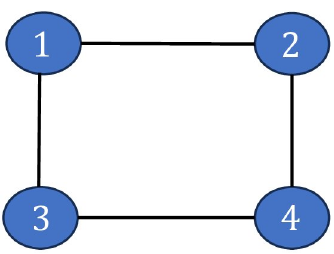}
    \caption{Topology of the oscillators.}
    \label{topology}
\end{figure}

\begin{figure}[htbp]
    \centering
    \includegraphics[width=0.85\linewidth]{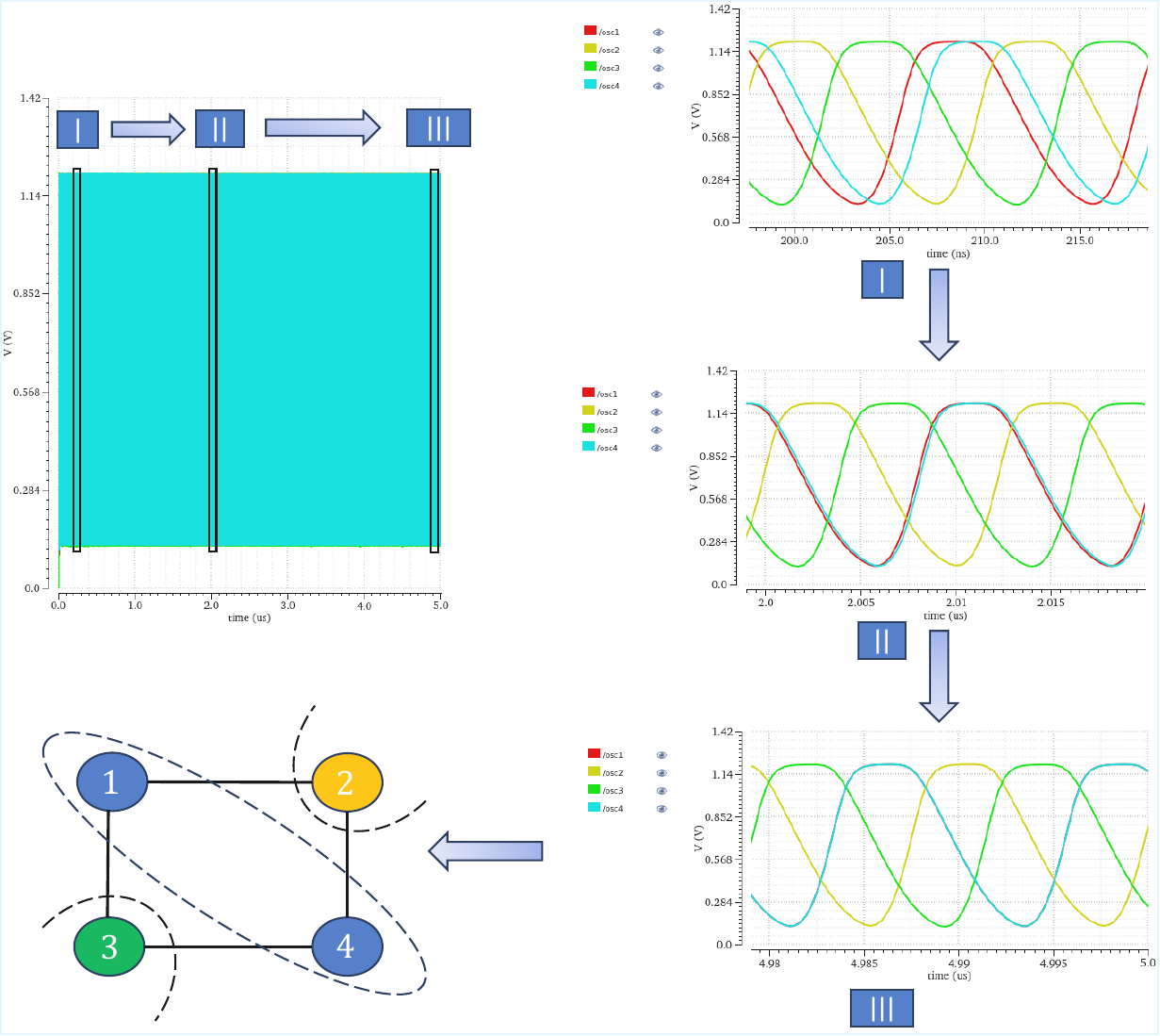}
    \caption{The evolution of waveforms of four ring oscillators without a $3$rd harmonic injection. The formation time of the three-state phase pattern is about $2{\rm \mu s}$, shown in the label 2.}
    \label{opm_waveform}
\end{figure}

\begin{figure}[htbp]
    \centering
    \includegraphics[width=0.75\linewidth]{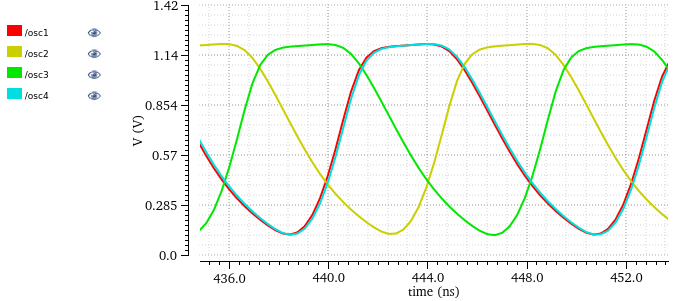}
    \caption{The evolution of waveforms of four ring oscillators with a $3$rd harmonic injection. The formation time of the three-state phase pattern is about $440{\rm ns}$.}
    \label{opm_waveform_sync}
\end{figure}

To demonstrate this mapping in hardware-realistic settings, we design a proof-of-concept circuit for a four-spin three-state OPM by using four coupled three-stage CMOS ring oscillators. The oscillators are interconnected according to the topology in Fig.~\ref{topology},
where the coupling weights $J_{ij} \in \{0,1\}$ and the target phase dynamics are
\begin{eqnarray*}
    \dot \theta_1 &=& \sin(\theta_1-\theta_2)+\sin(\theta_1-\theta_3) + \sin(2\theta_1-2\theta_2)+\sin(2\theta_1-2\theta_3) - K_{\rm s}\sin(3\theta_1), \nonumber \\
    \dot \theta_2 &=&\sin(\theta_2-\theta_1)+\sin(\theta_2-\theta_4) + \sin(2\theta_2-2\theta_1)+\sin(2\theta_2-2\theta_4) - K_{\rm s}\sin(3\theta_2), \nonumber \\
     \dot \theta_3 &=& \sin(\theta_3-\theta_1)+\sin(\theta_3-\theta_4) + \sin(2\theta_3-2\theta_1)+\sin(2\theta_3-2\theta_4) - K_{\rm s}\sin(3\theta_3),\nonumber \\
    \dot \theta_4 &=& \sin(\theta_4-\theta_3)+\sin(\theta_4-\theta_2) + \sin(2\theta_4-2\theta_3)+\sin(2\theta_4-2\theta_2) - K_{\rm s}\sin(3\theta_4).
\end{eqnarray*}
Each oscillator’s output is fed into two band-pass filters centered at $f$ and $2f$, and the filtered voltages are converted into current injections by voltage-controlled current sources (VCCSs) that implement the required amplitudes $A_1$, $A_2$ and delays $\tau_1$, $\tau_2$. Here we note that the fundamental frequency of each ring oscillator, in our case, is $f \approx 81{\rm MHZ}$. In the absence of a third-harmonic injection, time-domain simulation in Fig.~\ref{opm_waveform} shows that, starting from four distinct initial phases, the ring-oscillator network converges to three stable phase values, corresponding to the three Potts states. Specifically, the first and fourth oscillators converge to a common phase, forming three phase differences with the second oscillator and third oscillator. The phase differences between the oscillators are very close to the ideal values $0$, $\pm 2\pi/3$, and $\pm 4\pi/3$.
When a third-harmonic injection is introduced and the amplitude $A_{\rm s} =20 {\mu A}$ is applied, the formation time of the three-state phase pattern is reduced from about $2{\rm \mu s}$ to $440{\rm ns}$, shown in Fig.~\ref{opm_waveform_sync}, which demonstrates that the $q$th harmonic term not only creates the required multi-well potential but also accelerates convergence to the desired OPM attractors. The details of the derivation, the circuit diagram, and the parameter settings of the circuit are provided in the Supplementary Material.

\subsection{Sampling performance and combinatorial optimization benchmarks}\label{benchmarks}
\begin{figure}[htbp]
    \centering
    \includegraphics[width=0.85\linewidth]{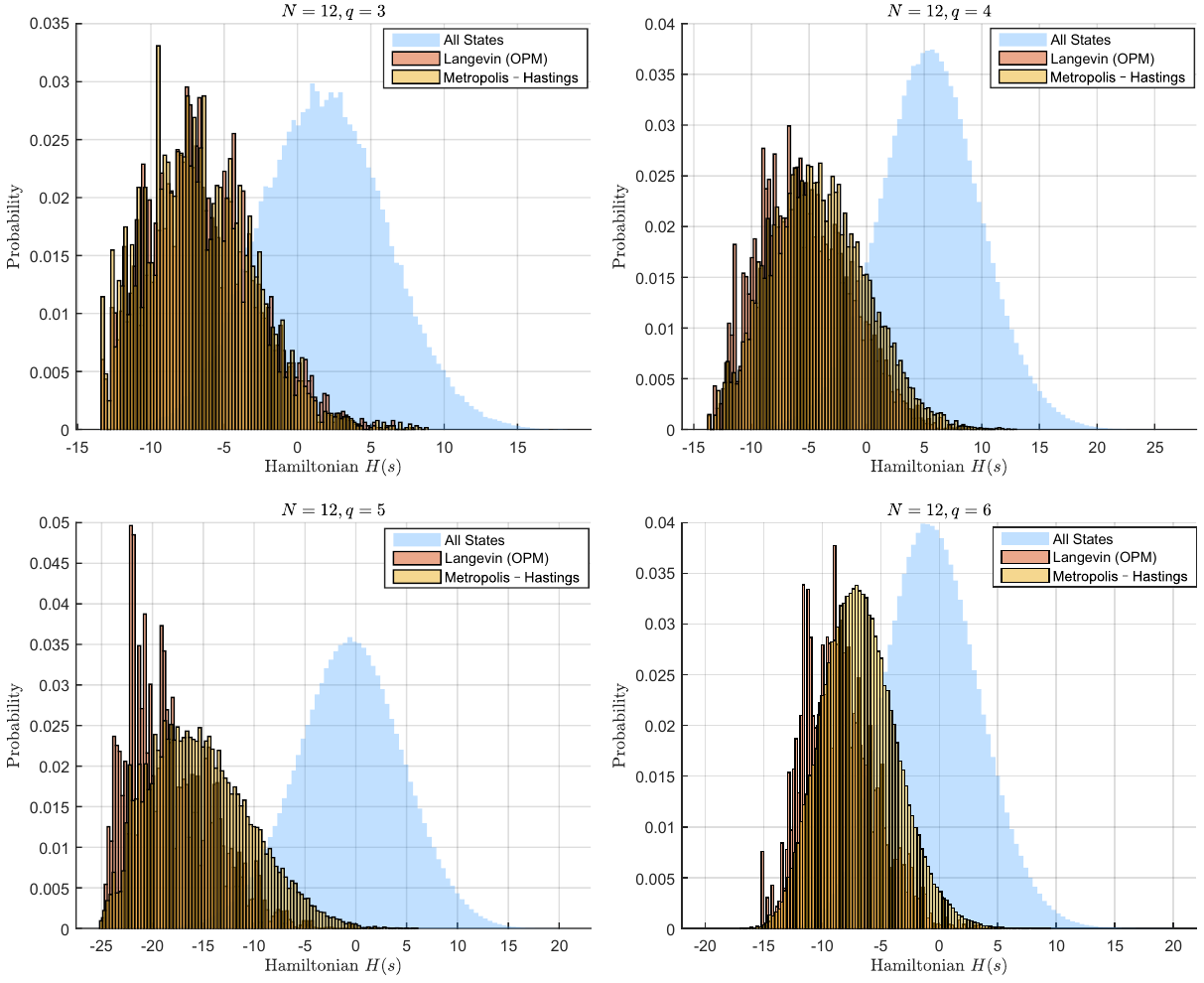}
    \caption{The distribution of samples: The reddish brown is the distribution of sample obtained from OPM. The yellow is the distribution of sample obtained from Metropolis-Hastings sampling. The light blue is the solution space.}
    \label{EX2.pdf}
\end{figure}

To assess the sampling quality of the OPM, we first consider small Potts models with $N=12$ spins and $q=3$, $4$, $5$, and $6$. For each value of $q$, we draw samples from the OPM dynamics \eqref{Lagevin}
and from a Metropolis–Hastings (MH) Markov chain targeting the exact Boltzmann distribution \eqref{boltzman} at inverse temperature $\beta =1$. The OPM trajectories are discretized by the quantization map introduced in Section \ref{lowenergybias}, and we keep the total number of OPM samples equal to the number of MH samples. The resulting energy histograms are shown in Fig.~\ref{EX2.pdf} for all four values of $q$.

As expected from the theoretical low-energy bias established in Section \ref{lowenergybias}, the OPM samples are clearly concentrated in the low-energy configurations. More importantly, the empirical energy distribution obtained from the OPM tracks closely that produced by MH sampling for all four values of $q$. Since MH sampling has the Boltzmann distribution as its stationary distribution, this agreement indicates that, after quantization, the OPM effectively samples from a distribution that is close to the Boltzmann distribution.

\begin{figure}[htbp]
    \centering
    \includegraphics[width=1\linewidth]{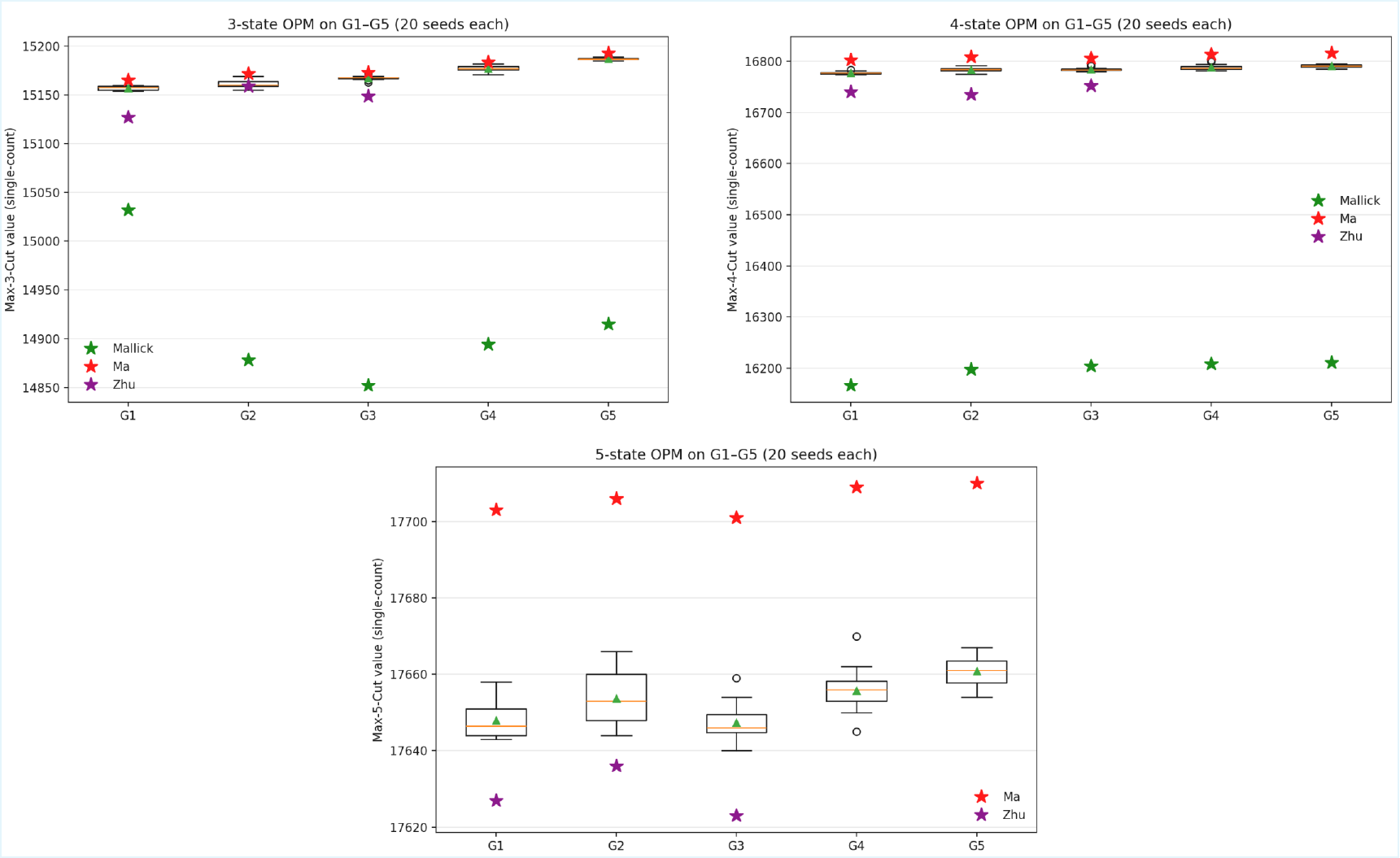}
    \caption{The benchmark on max-$K$-cut problems, $K=3$, $4$, and $5$.}
    \label{MaxKCutGest.pdf}
\end{figure}

We next test the OPM on large-scale combinatorial optimization problems by mapping max-$
K$-cut instances to Potts Hamiltonians. For the special case of the max-2-cut (Max-Cut) problem, the OPM reduces to an oscillator Ising machine, for which extensive benchmarks have already demonstrated excellent performance \cite{Wang:EECS-2020-12,chou2019analog,wang2019new}. Therefore, we do not further consider max-$2$-cut here and instead focus on max-$K$-cut problems for $K=3$, $4$, and $5$. Specifically, we consider the standard Gset graphs $\rm G1$-$\rm G5$ and study max-$3$-cut, max-$4$-cut, and max-$5$-cut, corresponding to $q=3$, $4$, and $5$, respectively. For each instance and each $K$, we run the OPM 20 times from independently randomized initial conditions and record the cut value obtained in each run. The resulting distributions of single-run performance are summarized as box plots in Fig.~\ref{MaxKCutGest.pdf}.
We compare the OPM against three representative algorithms. The first, Mallick et al. \cite{mallick2022computational}, is another physics-inspired dynamical system. The other two, proposed respectively by Ma et al. \cite{ma2017multiple} and Zhu et al. \cite{zhu2013max}, are more conventional heuristic methods that combine exhaustive search with sophisticated local search heuristics and, to the best of our knowledge, currently achieve the best known cut values on these benchmarks. In the figure, the colored stars indicate the best reported values from these three methods, while the OPM results appear as box plots. Across all graphs and for both max-$3$-cut and max-$4$-cut, the OPM consistently outperforms the physics-based method of Mallick et al. and the heuristic algorithm of Zhu  et al.. Note that even the worst OPM run yields a cut value higher than the best value reported by either of these approaches in almost cases. For max-$5$-cut, where results from Mallick  et al. are not available, the OPM again outperforms Zhu  et al.. When compared with Ma  et al., whose algorithm attains the best known cuts, the OPM’s performance is slightly but systematically lower. However, the gaps are very small, amounting to only a tiny fraction of a percent in cut value, and the best OPM runs lie very close to the best known results of Ma et al. on all instances. The best cut value among the four approaches is reported in Table~\ref{bestreport}.
\begin{table}[htbp]
\centering
\caption{The best reports on G1--G5.}
\setlength{\tabcolsep}{8pt}
\renewcommand{\arraystretch}{1.2}

\begin{tabular}{l *{12}{c}}
\toprule
& \multicolumn{4}{c}{Max-3-cut}
& \multicolumn{4}{c}{Max-4-cut}
& \multicolumn{4}{c}{Max-5-cut} \\
\cmidrule(lr){2-5}\cmidrule(lr){6-9}\cmidrule(lr){10-13}
& Ours & Mallick & Ma & Zhu
& Ours & Mallick & Ma & Zhu
& Ours & Mallick & Ma & Zhu \\
\midrule
G1 &  15,160 &  15,032 &  15,165 & 15,127 & 16,783 & 16,166 & 16,803 & 16,740 & 17,658 & -- & 17,703 &  17,627 \\
G2 &  15,169 &  14,878 &  15,172 & 15,159 & 16,792 & 16,197 & 16,809 & 16,735 & 17,666 & -- & 17,706 &  17,636\\
G3 &  15,172 &  14,852 &  15,173 & 15,149 & 16,794 & 16,204 & 16,806 & 16,752 & 17,659 & -- & 17,701 &  17,623 \\
G4 &  15,182 &  14,894 &  15,184 & --     & 16,802 & 16,208 & 16,814 & --     & 17,670 & -- & 17,709 &  --\\
G5 &  15,190 &  14,915 &  15,193 & --     & 16,795 & 16,211 & 16,816 & --     & 17,667 & -- & 17,710 &  --\\
\bottomrule
\end{tabular} \label{bestreport}
\end{table}
The benchmarks show that a straightforward implementation of the OPM as overdamped Langevin dynamics potentially delivers competitive performance on challenging large-scale max-$K$-cut problems.

\section{DISCUSSION}
\subsection{Structural minimality of the OPM Langevin dynamics}
\begin{figure}[htbp]
    \centering
    \includegraphics[width=0.7\linewidth]{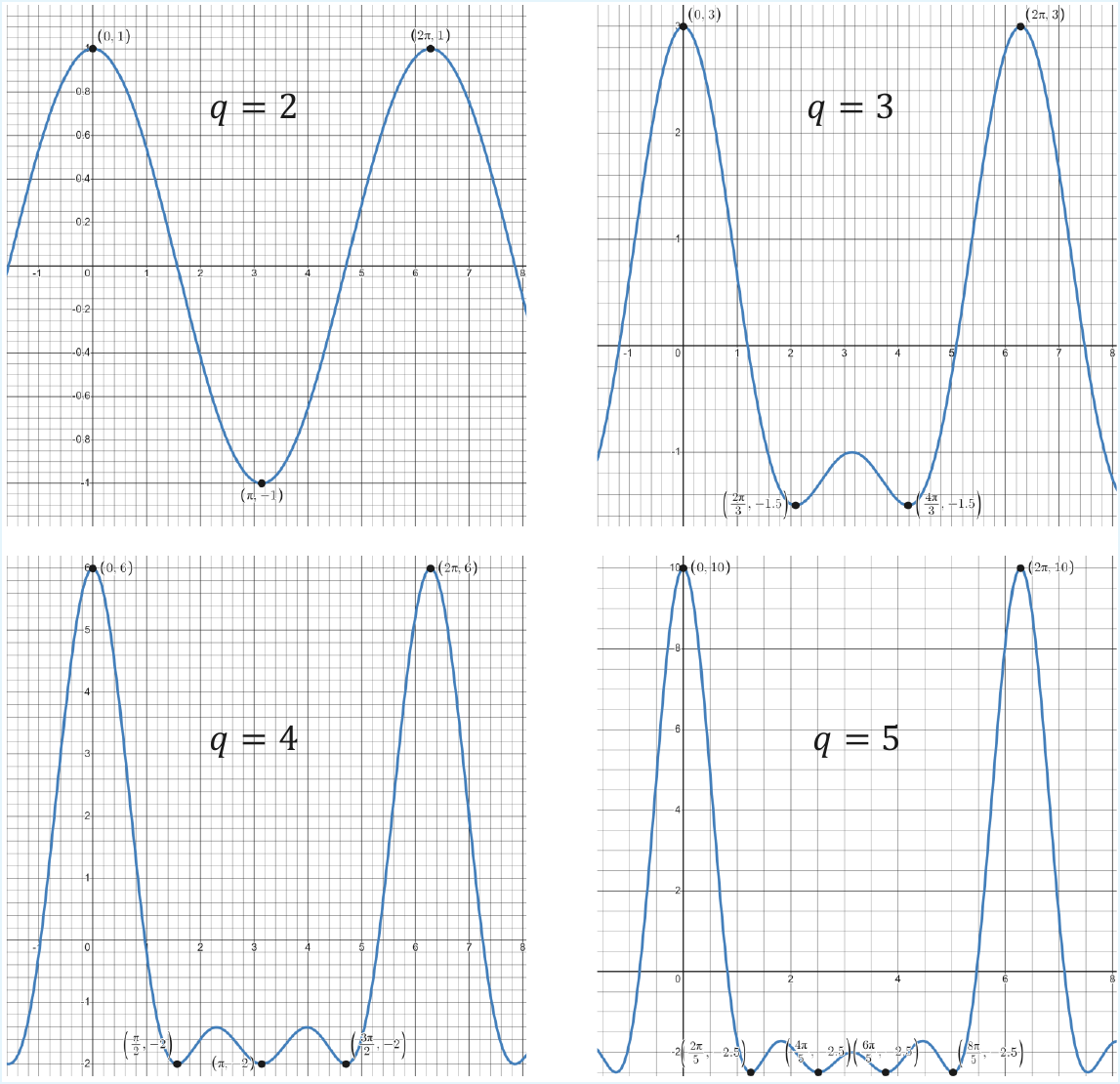}
    \caption{Interactive function for $q=2$, $3$, $4$, and $5$.}
    \label{minimalstructure}
\end{figure}

The $q$-state Potts Hamiltonian admits, at least in principle, infinitely many continuous relaxations and hence infinitely many associated overdamped Langevin dynamics. Even under the equilibrium-preserving constraint introduced in this work, namely, that every Potts configuration be mapped to an equilibrium of the continuous dynamics, and that the energy at these points relaxes the Potts Hamiltonian, there remains a large design space of possible energy functions and corresponding stochastic dynamics. In this context, a natural question is what distinguishes the specific OPM dynamics proposed here from other admissible Langevin realizations of the same Potts model. Our view is that the OPM occupies a corner of particular simplicity in this design space. It is helpful to view the OPM energy \eqref{UH} through the geometry of its pairwise interaction $f(x) = \sum_{m=1}^{q-1}((q-m)\cos(mx)),x=\theta_i-\theta_j$. The plots in Fig.~\ref{minimalstructure}, for $q=2$, $3$, $4$, and $5$, provide two pieces of physical intuition that jointly justify why this specific interaction is not just an equilibrium-preserving relaxation, but a structurally minimal one.
First, the phase differences $x = 2\pi k/q,k=0,1,\ldots,q-1$, appear as extrema of $f(x)$. When every $\theta_i$ lies on the $q$-grid, each edge contribution $J_{ij}f(\theta_i-\theta_j)$ sits at a stationary point, so the deterministic drift generated by $-\nabla U(\theta)$ vanishes on these configurations in a way that is robust to perturbations of the couplings $J_{ij}$. This is the sense in which the sampling points are structurally stable equilibrium points of the relaxed landscape. Second, the plots show a Potts-like degeneracy, that is, $f(0)>0$, whereas all nonzero grid differences $x= 2\pi k /q$ with $k=1,2,\ldots,q-1$ share the same negative value. Hence, the pairwise energy depends only on whether two spins are equal ($x=0$) or different ($x\ne 0$), and is insensitive to which different labels they take, which is exactly the equilibrium-preserving property needed for the relaxation to reproduce the Potts Hamiltonian on discrete states. While many alternative periodic interactions could be engineered to satisfy these two constraints (and trivial rescalings or shifts generate infinitely many variants), there is a sharp lower bound on the “structural complexity” measured by the number of extrema per period. Indeed, if a smooth $2\pi$-periodic function $g(x)$ makes every grid point $2\pi k /q$ a stationary point and enforces equal energy for all $k=1,2,\ldots,q-1$, by Rolle’s theorem each interval between them must contain at least one additional critical point. Thus, beyond the $q$
prescribed stationary points on the grid, at least $q-2$ extra extrema are unavoidable, implying a minimum of $2q-2$ critical points per period. The interaction $f(x)$ achieves precisely this limiting case (e.g., for $q=3$, one extra extreme is forced between $x=2\pi/3$ and $x= 4\pi/3$, and there are four extremes per period), explaining why the
OPM choice is the simplest interaction that simultaneously (i) pins all Potts sampling configurations as robust equilibrium points and (ii) collapses the relaxed energy on those equilibrium points to a Potts form.

\subsection{OPM as a framework for multi-valued associative memory}
Beyond its role as a sampler and optimization engine, the OPM naturally suggests a dynamical framework for associative memory. As shown in Section \ref{dynamicalpro}, in the deterministic form, each oscillator admits $q$ stable phases $\theta_i^\star = 2\pi k/q$, $k=1,2,\ldots,q-1$. These sampling points form a discrete set of attractors in phase space, with a one-to-one correspondence to Potts spin configurations. This structure is reminiscent of classical associative memory models, but is now generalized from binary neurons to multi-valued units with $q$ phase states. From this viewpoint, each sampling point can be regarded as a candidate memory pattern, and the associated basin of attraction quantifies the robustness of its retrieval. The structural stability of the sampling points under perturbations of the couplings $J_{ij}$ and parameters $K$ and $K_{\rm s}$ implies that these patterns are not fine-tuned, that is, any perturbation of the couplings and the parameters does not destroy the existence of the attractors, but only deforms their basins. Dynamic stability thus becomes the operational criterion for whether a desired Potts pattern is reliably stored in a given OPM instance; namely, a pattern is stored if its corresponding sampling point is asymptotically stable. This perspective raises a series of questions that parallel, but do not trivially reduce to, those studied in Hopfield-type networks and oscillatory associative-memory models. In particular, it is natural to ask how many distinct Potts patterns can be embedded into an OPM such that each remains dynamically stable with a macroscopic basin of attraction, and how this capacity scales with the number of oscillators $N$ and the number of states $q$. A recent work \cite{11186164} has investigated several basic properties of the oscillator Ising machine ($q=2$) as an associative-memory network. Our discussion places this special case into a broader dynamical perspective, suggesting OPMs as a systematic route to multi-valued associative memory when $q\ge 3$.

We therefore view the OPM as a promising starting point for a systematic theory of multi-valued associative memory in oscillator networks. Key open problems include characterizing the scaling of the number of stably retrievable Potts patterns as a function of $N$ and
$q$, understanding how coupling weights and parameters affect retrieval basins, and comparing the resulting capacity to that of classical Hopfield and Potts–Hopfield networks. Addressing these questions would not only clarify the memory-theoretic implications of the OPM construction, but also offer a dynamical-systems perspective on how discrete, multi-valued memories might be embedded in oscillatory neural activity.

\section{Materials and Methods}
Circuit-level simulations were performed using Cadence Virtuoso. The complete circuit schematics, transistor-level parameters (e.g., device dimensions and supply voltages), and simulation setups (e.g., transient settings and measurement configurations) are provided in the Supplementary Material.

The simulation results detailed in Fig.~\ref{EX2.pdf} pertain to networks of $N=12$ and the number of states $q\in \{3,4,5,6\}$.  The graph is randomly generated. In the present manuscript, we use a fixed set of parameters $K \in \{1,3/8,4/25,,1/12\}$ for $q\in \{3,4,5,6\}$, respectively. We use the standard Euler-Maruyama scheme for the OPM dynamics \eqref{OPMdyn}, where we set the step size $\epsilon = 10^{-3}$ and fix the inverse temperature $\beta = 1$ and  $K_{\rm s} =20$. The coupling weights $J_{ij}$ are randomly chosen. In the quantization process, we set $a = 0.3\pi/q$ for $q=3$, $4$, $5$, and $6$. In addition, we add Metropolis-Hastings sampling for comparison. We keep the number of sampling points obtained by the OPM and that of samples obtained by Metropolis-Hastings sampling the same. All simulations in this part are implemented in MATLAB.

Across all Max-$K$-cut experiments, we integrate the phase SDE for a total time $T_{\rm end} = 250$ with step size ${\rm d} t = 10^{-3}$ using the Euler–Maruyama method, and simulate $B = 300$ trajectories in parallel. During the run, we anneal only the noise amplitude $\sigma(t) : =\sqrt{2\beta^{-1}}$ and the strength $K_{\rm s}(t)$, while keeping the coupling $K$ fixed. The annealing schedule uses a cosine-type interpolation from the specified start to end values. Specifically, for max-$3$-cut, we use $K = 15$, $\sigma(t): = 8 \to 0.4$, $K_{\rm s}(t): 15 \to 450$. For max-$4$-cut, we use $K = 5$, $\sigma(t): 8 \to 0.4$, $K_{\rm s}(t): 5 \to 120$. For max-5-cut, we use $K = 1.25$, $\sigma(t): 5 \to 0.3$, $K_{\rm s}(t):1.25 \to 60$. Candidate discrete solutions are evaluated using a sparse-to-dense evaluation strategy with a base evaluation period $2000$ steps, tightened after a no-improvement window of $250$ steps for a span of $120$ steps. All runs are executed in PyTorch with GPU acceleration on an NVIDIA GeForce RTX 2080 Ti.

\textbf{Data and materials availability}: There are no data underlying this work. The code used for the max-$K$-cut simulations and figure generation is available at https://doi.org/10.5281/zenodo.17989975.

\nocite{*}

\bibliography{apssamp}

\end{document}